\documentclass[twocolumn,showpacs,preprintnumbers,prb,amsmath,
amssymb,superscriptaddress]{revtex4}
\usepackage{graphicx}
\usepackage{dcolumn}
\usepackage{epsfig}

\begin{document}

\title{Geometrical enhancement of the proximity effect in quantum wires\\ 
with extended superconducting tunnel contacts}

\author{Giorgos Fagas}
\affiliation{Tyndall National Institute, Lee Maltings, Prospect Row,
Cork, Ireland}
\email{gfagas@tyndall.ie}
\author{Grygoriy Tkachov}
\affiliation{
Institut f\"{u}r Theoretische Physik, Universit\"{a}t
Regensburg, 93040 Regensburg, Germany}
\email{Grigory.Tkachov@physik.uni-r.de}
\author{Andreas Pfund}
\affiliation{
Institut f\"{u}r Theoretische Physik, Universit\"{a}t
Regensburg, 93040 Regensburg, Germany}
\author{Klaus Richter}
\affiliation{
Institut f\"{u}r Theoretische Physik, Universit\"{a}t
Regensburg, 93040 Regensburg, Germany}

\date{\today}

\begin{abstract}
We study Andreev reflection in a ballistic one-dimensional channel 
coupled in parallel to a superconductor via a tunnel barrier of finite length $L$. 
The dependence of the low-energy Andreev reflection probability $R_A$ on $L$ 
reveals the existence of a characteristic length scale $\xi_N$ beyond
which $R_A(L)$ is enhanced up to unity despite the low interfacial transparency. 
The Andreev reflection enhancement is due to the strong mixing of particle and
hole states that builds up in contacts exceeding the coherence length $\xi_N$,
leading  to a small energy gap (minigap) in the density of states of the normal
system. The role of the geometry of such hybrid contacts is discussed  
in the context of the experimental observation of zero-bias Andreev anomalies
in the resistance of extended carbon nanotube/superconductor junctions 
in field effect transistor setups. 
\end{abstract}


\maketitle

\section{Introduction}
\label{sec:S1}

The interest in proximity-induced superconductivity 
in one-dimensional (1D) electron systems~\cite{Fazio,Maslov,Takane,Affleck,Wei,Titov,Gonzalez,Vish,Nikolic,Lee,Jiang,Vecino,Tanaka} 
has recently revived in the light of successful experiments
on electron transport through carbon nanotubes contacted by
superconductors~\cite{Kasumov,Morpurgo,Schoenen,Haruyama}. 
Despite possibly strong electron-electron interactions~\cite{Schoenen,Bockrath},
in many situations transport properties
of metallic single-walled carbon nanotubes can be interpreted within a
ballistic model assuming two conduction bands at the Fermi level~\cite{Dres}. 
Hence, superconductor/carbon nanotube (S/CN) junctions can to some extent be viewed as 
an experimentally accessible case of 1D ballistic proximity structures~\cite{Morpurgo}.

Like in conventional normal metal/superconductor (N/S) junctions, the extent
to which the proximity effect modifies the electronic properties of carbon
nanotubes strongly depends  on the quality of S/CN interfaces. 
In S/CN/S junctions with highly nontrivial end-bonding of the tubes 
it is possible to achieve high transparency contacts and observe induced
supercurrents between the S banks~\cite{Kasumov,Schoenen,Haruyama}.
In a more conventional field-effect transistor setup a superconductor
is sputtered on top of a nanotube covering it from the ends 
and in this way connecting it to the leads~\cite{Morpurgo}.
Such contacts exhibit no observable superconducting coupling,
probably because of a Schottky barrier formed at the S/CN interfaces.
Nevertheless, in this case the proximity effect manifests itself 
as a pronounced zero-bias {\em dip} in the low-temperature resistance
to which either of the S/CN interfaces contributes independently~\cite{Morpurgo}.

The sensitivity of the zero-bias resistance anomaly to the temperature~\cite{Morpurgo,Haruyama}
suggests that it can be attributed to the conversion of a normal current into a supercurrent via
the Andreev reflection process~\cite{Andreev} during which particles with
energies much smaller than the superconducting gap  $\Delta$ are coherently
scattered from an S/CN interface as Fermi sea holes back to the normal system.
Under assumption of each of the S/CN interfaces acting independently~\cite{Morpurgo} 
and in the picture of non-interacting electrons,
such an interpretation must reconcile with the well-established 1D scattering model
for a single N/S contact~\cite{BTK}. However, for a point contact of average quality 
(between metallic and tunnel regimes) the theory of Ref.~\onlinecite{BTK} predicts
a zero-bias resistance {\em peak} at temperatures $T<\Delta/k_B$, that 
is exactly the opposite to the experimental findings of Refs.~\onlinecite{Morpurgo,Haruyama} 
in the same temperature regime.

Deviations of Andreev reflection physics in 1D proximity structures from the standard model of 
Ref.~\onlinecite{BTK} have so far been ascribed 
to repulsive electron interactions~\cite{Maslov,Takane,Affleck,Vish,Lee} or 
disorder in the normal channel~\cite{Tanaka}.  
In the present paper we show that clean non-interacting 1D systems 
can also exhibit unusual Andreev reflection properties 
if the contact to the superconductor is not a point-like one.  
Such contacts naturally occur in field effect 
transistor setups due to a finite overlap between a nanotube and a superconductor      
coupled in parallel. In particular, in the device of Ref.~\onlinecite{Morpurgo}
this overlap was as large as $1 \mu m$.
To demonstrate the importance of the contact geometry, 
we develop a scattering
model for phase-coherent electron transport through a normal 1D ballistic channel
part of which is in parallel coupling to a 2D superconductor via a low-transparency
barrier (Fig.~\ref{fig:1}). This model is in many aspects different from
the device of Ref.~\onlinecite{Morpurgo} and it is not expected to desribe
all the experimental features. However, it captures the most essential,
for our purposes, attribute of the S/CN contacts, namely their extended character. 
Moreover, the proposed geometry may serve as a minimum model accounting for the
zero-bias resistance features reported in Refs.~\onlinecite{Kroemer,Rahman,Weiss} 
for extended planar contacts between
ballistic 2D electron systems and superconductors,
whose cross-sectional structure is similar to that shown in Fig.~\ref{fig:1}.

Our numerical simulation of elastic quasiparticle scattering shows that the
probability of Andreev reflection depends on the length $L$ of 
the contact, approaching unity as $L$ exceeds a certain length scale $\xi_N$
larger than the coherence length in the superconductor $\xi_S$.
Most importantly, at zero energy high-probability Andreev reflection 
occurs {\it at any finite interfacial transparency} for sufficiently long
contacts. This is in sharp contrast to the situation in point
junctions~\cite{BTK}. To rationalize this result we perform a
numerical analysis of the quasiparticle density of states (DOS) in the region of
the 1D system coupled to the superconductor in the limit $L\gg\xi_N$.
The DOS is found to have a proximity-induced gap (minigap) at the Fermi level
whose  size $E_g\approx (\xi_S/\xi_N)\Delta $ is much smaller than the gap
$\Delta$ in the superconductor. The minigap $E_g$ scales with the interfacial
transparency, which implies that it is due to the formation of mixed particle-hole
(Andreev) states~\cite{MiniTh}. A comprehensive analysis of the energy
dependence of electron scattering reveals that the gapped excitation spectrum 
in the proximity region results in the enhancement of the Andreev reflection
probability $R_A(\epsilon)$ at finite (but small) energies
$\epsilon <E_g\ll \Delta$ followed by its decrease at 
intermediate energies $E_g<\epsilon<\Delta$. At the edge of the superconducting gap 
($\epsilon =\Delta$) the dependence $R_A(\epsilon)$ exhibits one more peak
typical for tunnel junctions~\cite{BTK}. These features dominate the bias
voltage dependence of the differential resistance which at $T\sim E_g/k_B$
displays a dip around the zero voltage similar to that observed
in Ref.~\onlinecite{Morpurgo}.

Previously, zero-bias conductance anomalies have been extensively studied
in mesoscopic superconducting contacts with
semiconductors
~\cite{Cast,vanWees}$^{,}$\cite{Kroemer,Green,Been,Marmorkos,Nitta,Poirier,Lambert,Les}$^{,}$
\cite{Rahman,Imry,Weiss,Batov,Ferry}
and metals~\cite{Petrashov,Pannetier,Golubov,Belzig,Kad,Altland,Anthore,Tanaka}.
These studies have predominantly focused on the diffusive transport regime.
According to the semiclassical scattering interpretation of
Ref.~\onlinecite{vanWees}, the excess conductance (i.e. exceeding the value
predicted by the theory of Ref.~\onlinecite{BTK}) is a signature of the correlated
particle-hole motion arising from multiple Andreev
reflections at the interface mediated by elastic scatterers in the normal system. 
Even for a low-transparency contact the cumulative  Andreev reflection probability
can be $\sim 1$ for trajectories hitting  the interface many times provided
that the area of the contact is sufficiently large. In the less explored regime
of ballistic propagation, a similar process, called sometimes reflectionless
tunneling~\cite{Been}, occurs in ballistic quantum wells in parallel long
contacts with superconductors. In these setups, the multiple Andreev
reflections are due to the back wall of the quantum well~\cite{Kroemer,MiniTh,Imry}.

Therefore, the low-bias excess conductance discussed in the present paper 
is a pronounced case of the reflectionless tunneling in ballistic systems
where $R_A(\epsilon)$ can be interpreted as the cumulative Andreev reflection
probability due to the correlated quantum particle-hole motion in the region of
the 1D channel coupled to the superconductor. It is also known that such
correlations can lead to a minigap in the quasiparticle
DOS~\cite{MiniTh,Golubov,Belzig,Altland,Billiards,Ihra,Cserti}.
We note that the previous studies of reflectionless transport
Refs.~\onlinecite{vanWees} and \onlinecite{Imry} dealt with multiple Andreev
reflections semiclassically and at small energies $|\epsilon| \ll E_g$.
Our quantum scattering approach is capable of describing the entire energy
dependence of the subgap conductance which shows the crossover from
reflectionless tunneling to independent electron tunneling through the barrier
at the N/S interface.
Besides, our numerical technique allows us to tackle the realistic geometry of
finite-length parallel N/S contacts and to obtain an accurate complete
dependence $R_A(L)$ which has not been studied in the previous models.

The structure of the article is as follows. In Section~\ref{sec:S2},
after a brief description of our system, we present the numerical results
for the DOS in the 1D channel. An analytical model is also developed that
helps to rationalize the low-energy regime. The length and energy dependence
of the Andreev reflection probability is analyzed in Section~\ref{sec:S3}.
In Section \ref{sec:S4}, we summarize the implications of our results with a
concluding discussion on the bias voltage dependence of the resistance.

\section{Two-gap spectral properties of extended superconducting 
tunnel contacts} 
\label{sec:S2}

In this section we study the density of states (DOS) in a 
quasi-one-dimensional electron system  (Q1DES) coupled in
parallel to a superconducting film via an interfacial barrier. 
We consider the two geometries shown in Figs.~\ref{fig:1}(a) and (b).
The heterostructures are assumed two-dimensional and located
in the plane $x,z$. Our results can be easily extended to 
an out-of-plane periodic structure defining a
quasi-two-dimensional electron system on the normal side.

\subsection{Description of the method}
\label{sec:S2A}

\begin{figure}[t]
\centerline{
\epsfig{figure=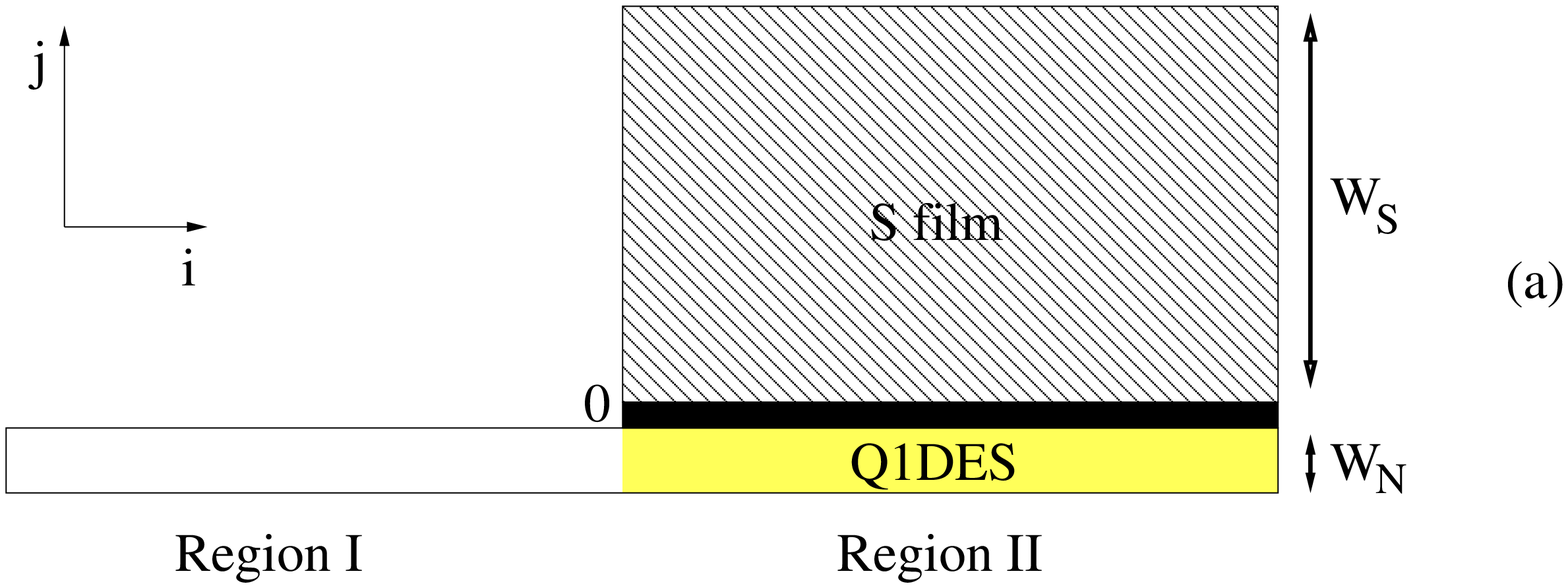, width=.9 \hsize ,height=.45 \hsize}
}
\centerline{
\epsfig{figure=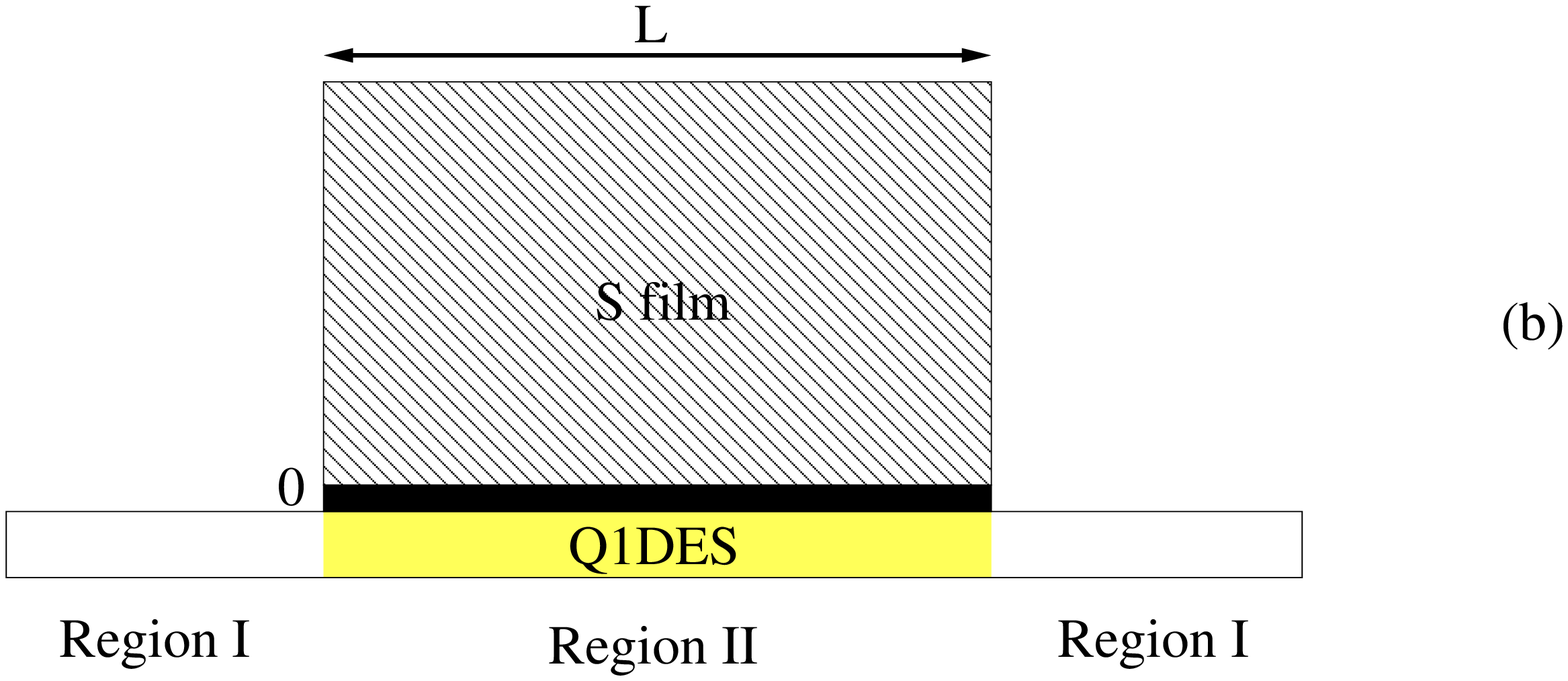, width=.9 \hsize ,height=.45 \hsize}
}
\caption{
Quasi-one-dimensional electron system (Q1DES) coupled to: (a) semi-infinite 
and (b) finite length $L$ superconductor contacts. The interfacial
barrier is depicted in black. The Q1DES width is chosen to be $W_N\approx\lambda_F/2$,
where $\lambda_F$ is the Fermi wavelength.}
\label{fig:1}
\end{figure}

To analyze the superconducting proximity effect in the Q1DES we employ
a numerical approach to solve the Bogolubov-de Gennes (BdG) equation
\begin{eqnarray}
\left(
\begin{array}{cc}
\hat H
& \hat \Delta \\
\hat \Delta
& - \hat H^\ast
\end{array}
\right)
\left(
\begin{array}{cc}
u(x,z)\\
v(x,z)
\end{array}
\right)=
\epsilon
\left(
\begin{array}{cc}
u(x,z)\\
v(x,z)
\end{array}
\right)
\label{Neq1}
\end{eqnarray}
for the electron $u(x,z)$ and hole $v(x,z)$ wavefunctions.
The method allows for performing a straight-forward discretisation
on a real-space grid of the one-particle Hamiltonian
$\hat H= -(\hbar^2/2m) (\partial_x^2 + \partial_z^2) + U(x,z) - \mu$,
the pairing potential $\hat \Delta= \Delta(x,z)$,
and the potential $U(x,z)$ (to  be defined later).
$\mu$ and $m$ are the chemical potential and the electron mass, both
constant throughout the entire system.
No translational invariance in the $x$-direction is invoked so that
studies of the structures in Figs.~\ref{fig:1}(a) and (b) are possible.

The discretized BdG equations read
\begin{eqnarray}
&
\left[ \epsilon-(4\gamma+U(i,j)-\mu) \right] u(i,j)
\nonumber\\
&
+\gamma \sum_{i^\prime j^\prime} u(i^\prime,j^\prime)
- \Delta(i,j)v(i,j)=0,&
\nonumber\\
&
\label{Neq2}&\\
&
\left[ -\epsilon-(4\gamma+U(i,j)-\mu) \right] v(i,j)
\nonumber\\
&
+\gamma \sum_{i^\prime j^\prime} v(i^\prime,j^\prime)
+ \Delta(i,j)u(i,j)=0,&
\nonumber
\end{eqnarray}
where $i$ and $j$ refer to sites on a two-dimensional lattice 
in the $x$ and $z$ directions, respectively, and
primes denote summation over nearest neighbours.
The origin of the coordinate frame is
indicated in Figs.~\ref{fig:1}(a) and~\ref{fig:1}(b) by the zero.
If required, Eq.~(\ref{Neq2}) can be generalized to a position dependent
effective mass for specific materials~\cite{PRB84MB}.

The parameters of the numerical scheme are as follows. 
The potential $U(i,j)$
is infinite everywhere outside the N and S systems. For every $j$ 
within the materials,
$U(i,j)=0$ for $i < 0$ and $U(i,j)=U_o\ge 0$ otherwise. 
A positive potential step $U_o$ accounts for the fact that the coupling
to the superconductor may result in a slight reduction of 
the Fermi energy of the Q1DES in region (II)  
in Figs.~\ref{fig:1}(a) and~\ref{fig:1}(b) compared to that in the 
uncoupled region 
(I) (cf Ref.~\onlinecite{Weiss}). This turns out to be important 
when considering scattering of quasiparticles incident at region (II), 
which is 
analyzed in the next section.  
The pairing potential $\Delta(i,j)$ in Eq.~(\ref{Neq2})
is assumed position-independent and equal to $\Delta\cdot\delta_{ij}$ (s-wave)
in the superconductor and zero everywhere else.
Although the self-consistency is ignored,
the stepwise order parameter has proved to be a satisfactory approximation for 
studying the proximity effect in clean systems~\cite{Been,Lambert}.
The absolute value of $\gamma$ is inverse proportional to the mesh 
parameter
$\alpha$, which is varied until convergence of the results is reached.
A sufficient condition is $\xi_S, \lambda_F \gg \alpha$, where
$\xi_S=\hbar v_F / 2\Delta$ is the superconducting coherence 
length
and $\lambda_F$ ($v_F$) is the Fermi wavelength (velocity).
To simulate the effect of a relatively thick superconducting film, we 
consider $W_S/\xi_S = 15$ for the spectral properties and
$W_S/\xi_S = 50$ for the results of Section~\ref{sec:S3}. Increasing this
ratio does not have any quantitative effect at $|\epsilon| < \Delta$.
In particular, all features discussed below are already observed for
$W_S/\xi_S \approx 3$ but with prominent finite-size effects for 
high quasiparticle energies $|\epsilon| > \Delta$.
The width of the normal region is fixed to $W_N/\lambda_F \approx 1/2$,
allowing only one propagating mode along the Q1DES.
The ratio $\xi_S/\lambda_F$ is chosen to be 2 
(see also endnote~\onlinecite{ratio}).

A tunneling barrier at the N/S interface (dark area in Fig.~\ref{fig:1})
is introduced via an effective Hamiltonian equivalent to adding
\begin{eqnarray}
&
\sum_{i^\prime j^\prime}
\left( \gamma_{NS}-\gamma \right)
(\delta_{j,0} \delta_{j^\prime,1} + \delta_{j,1} \delta_{j^\prime,0} )\; 
u(i^\prime,j^\prime) = 0,&
\nonumber\\
&
\label{Neq3}&\\
&
\sum_{i^\prime j^\prime}
\left( \gamma_{NS}-\gamma \right)
(\delta_{j,0} \delta_{j^\prime,1} + \delta_{j,1} \delta_{j^\prime,0} )\;
v(i^\prime,j^\prime) = 0,&
\nonumber
\end{eqnarray}
to Eq.~(\ref{Neq2}) for every $i$ within region (II).
Essentially, the above boundary conditions define the coupling
between the normal and the superconducting systems via the interfacial
constant $\gamma_{NS}$. The latter may arise from a formal procedure~\cite{JMP62L}
that  projects out the degrees of freedom within an insulating layer with a very
high barrier, when neglecting the energy and momentum dependence of the
penetration length. In what follows, we express all energies in units of
$\gamma$ for convenience.

\begin{figure}[t]
\centerline{
\epsfig{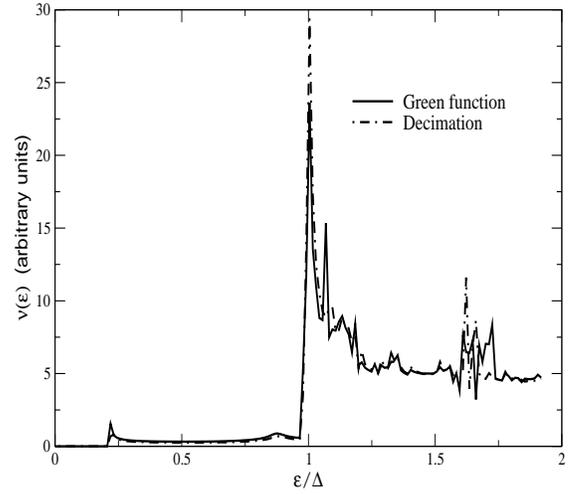}
}
\caption{
DOS $\nu(\epsilon)$ of a ballistic Q1DES-superconductor system. The solid line 
corresponds to the semi-infinite geometry shown in Fig. \ref{fig:1}(a)
for which $\nu(\epsilon)$ is calculated directly from the Green 
function of the system. The dashed line corresponds to the case of a
finite (but relatively long $L/\xi_S \gg 1$) S film   
[Fig. \ref{fig:1}(b)] where we use the decimation technique.}
\label{fig:2}
\end{figure}
\vspace{1cm}
\begin{figure}[t]
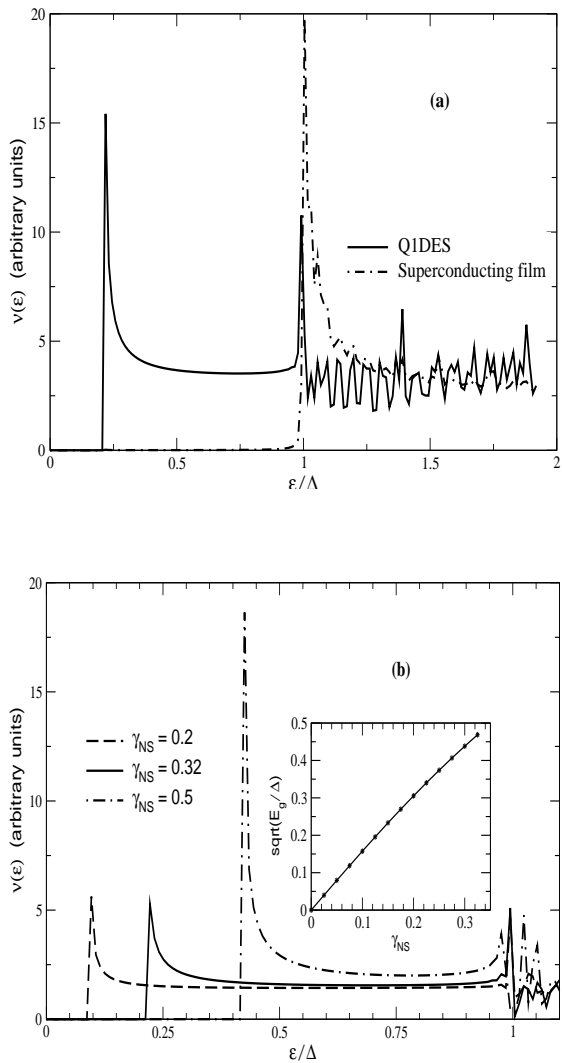

\centerline{
\epsfig{figure=fig3a.eps, width=.85 \hsize ,height=.75 \hsize}
}
\vspace{1.05cm}
\centerline{
\epsfig{figure=fig3b.eps, width=.85 \hsize ,height=.75 \hsize}
}
\caption{(a) Decomposition of the DOS $\nu(\epsilon)$ to that arising from the Q1DES 
(solid curve)
and the superconductor (dashed curve). Both curves are scaled to 
compare with each other. 
(b) DOS in the proximity region for various interfacial 
transparencies. Inset: the minigap $E_g$ depends
quadratically on the interfacial coupling $\gamma_{NS}$
at low transparencies, and starts deviating at higher values
of $\gamma_{NS}$.}
\label{fig:3}
\end{figure}

To calculate the DOS of the hybrid N/S system (region (II) in
Fig.~\ref{fig:1}) and, later, to study the scattering and transport properties
of quasiparticles incoming from normal region (I), we use a volatile numerical
method used in studies of the magnetoresistance of hybrid
systems~\cite{PRB99TSL,PRB99SLJ}, phonon transport~\cite{PRB04FKL},
and more recently in molecular electronics~\cite{CPL04FKE}.
It is based on recursive Green function techniques. Although some of 
the implementation details may differ \cite{TIMES}, the main stages of the
computational scheme are explained in Ref. \onlinecite{PRB99SLJ}.

\subsection{Quasiparticle density of states}
\label{sec:S2B}

The proximity effect is reflected in the DOS, $\nu (\epsilon)$, of the 
hybrid system plotted in Fig.~\ref{fig:2} for $\gamma_{NS}=0.32$ and $U_o=0$.
The solid line corresponds to the semi-infinite geometry shown 
in Fig. \ref{fig:1}(a). In this case $\nu (\epsilon)$
is calculated from the Green function
$G(i,j,i^\prime,j^\prime;\epsilon)$ of region (II)
via the well-known relation
$\nu (\epsilon)=-\frac {1}{\pi N} \Im \sum_{i,j} G(i,j,i,j; \epsilon)$,
where the summation is over all lattice sites in the hybrid part of the junction;
the factor $N$ normalises the area under the curve to a reference unit
and $\Im$ means the imaginary part. The dashed line corresponds to the geometry of 
Fig.~\ref{fig:1}(b) for a relatively long wire $L/\xi_S \gg 1$.
In this case, we obtain $\nu(\epsilon)$ using 
a recursive technique (negative-factor counting \cite{Mark}) 
that allows one to calculate the effective (renormalised) interaction 
between the normal leads~\cite{JMP62L} 
by projecting out the degrees of freedom of the middle region (II).

Both approaches reveal the formation of a two-gap structure: a smaller 
gap (minigap) at $E_g\approx 0.21\Delta$ and the usual BCS singularity at $\epsilon 
=\Delta$ with a finite quasiparticle contribution at intermediate energies, 
$E_g<\epsilon <\Delta$. 
Figure \ref{fig:3}(a) shows the DOS separately for the Q1DES (solid 
line) and for the superconductor (dashed line) from which one can conclude that 
the smaller gap opens in the DOS of the Q1DES. This observation along 
with the dependence of $E_g$ on the coupling to the superconductor shown
in Fig.~\ref{fig:3}(b) suggests that the minigap formation is
a signature of the superconducting correlations induced 
in the Q1DES. They are maintained in the course of multiple Andreev 
reflections in the channel which mix particle and hole
states with energies below the effective pairing energy coinciding 
with $E_g$~\cite{MiniTh}. At higher energies
$E_g\ll\epsilon <\Delta$ the electrons and holes in the Q1DES are 
weakly correlated and hence can be treated as one-particle excitations. 

In mesoscopically large diffusive N/S systems the formation of the minigap 
has been studied in a number of theoretical papers
(see, e.g, Refs.~\onlinecite{Golubov,Belzig,Altland}). In the clean limit,
the minigap structure has been analyzed to some extent in
billiard geometries resembling quantum dots~\cite{Billiards,Ihra,Cserti}.

Below we develop an analytical model, close in spirit to our numerical 
approach, that provides a simple description of the superconducting
correlations in ballistic wires based on a 1D BdG-like equation with an
{\it effective proximity-induced} pairing energy $E_g$.

\subsection{Proximity effect in a clean quantum wire: 
An analytical model}
\label{sec:S2C}

Although the physical mechanism responsible for the minigap formation 
in ballistic 2D electron systems has been explored in Ref.~\onlinecite{MiniTh}, 
the proposed method of derivation of $E_g$ heavily relies 
on the following assumptions. Firstly, the pairing potential
in the superconductor $\Delta(z)$ was assumed homogeneous. Secondly,
a finite-thickness normal system was modelled by 
a rectangular confining potential $U_c(z)$ enabling a plane wave 
description of the multiple reflection in the normal channel. 
The advantage of the model of Ref.~\onlinecite{MiniTh} is that it allows one to
obtain the minigap for an arbitrary interfacial transparency.
In this subsection we present an alternative microscopic derivation of 
$E_g$ that does not rely on any particular models for $\Delta(z)$ and 
$U_c(z)$, but is restricted to low interfacial transparencies
and low energies $|\epsilon|\sim E_g\ll\Delta$. 
By focusing on this case (weak-coupling regime)
we would like to emphasize that the effects related to the minigap 
formation can be observed even in samples with average interfacial quality 
provided that the temperature is low enough.  
As in Ref.~\onlinecite{MiniTh}, we also assume the translational 
invariance along the N/S interface and 
different Fermi energies $\mu_S \gg \mu_N$ and Fermi momenta $p_S \gg 
p_N$ on the S and N sides. 

It is convenient to rewrite the BdG equation (\ref{Neq1})
for the two-component wavefunction 
$\psi_p(z)=[u_p(z),v_p(z)]^T$ in the superconductor ($z\geq 0$) 
in a more compact form:
\begin{eqnarray}
\left[
\epsilon\sigma_3+\left(E_S+\frac{\hbar^2}{2m}
\partial_z^2 \right)\sigma_0-\Delta(z)i\sigma_2
\right]\psi_p(z)=0.\label{EqS}
\end{eqnarray}
Here, $p\equiv p_x$ is the momentum parallel to the interface; 
$\sigma_{2,3}$ and $\sigma_0$ are the Pauli and unity 
matrices, respectively. In the normal system ($z\leq 0$) the equation for 
$\psi_p(z)$ is
\begin{eqnarray}
&\left[
\epsilon\sigma_3+
\left(E_N+\frac{\hbar^2}{2m}
\partial_z^2-U_c(z)\right)\sigma_0\right]\psi_p(z)=0,\quad&
\label{EqN}
\end{eqnarray}
with $E_{N,S}=\mu_{N,S}-p^2/2m$.
The confining potential $U_c(z)$ defines a Q1D channel 
with a localized electron wavefunction $\phi(z)$
in the $z$-direction. 

The interfacial barrier is assumed rectangular with the electron 
penetration length  $\kappa^{-1}_0=\hbar/(2mU)^{1/2}$ determined by 
the barrier height $U$ measured from the Fermi energy. 
Inside a high enough barrier one can 
neglect the energy and momentum dependence of the penetration 
length and write the BdG 
equation as
$
[\partial^2_z-\kappa^2_0]{\tilde\psi}_p(z)=0,\quad 0\leq z\leq a,
$
where $a$ is the barrier thickness.
We introduce a special notation ${\tilde\psi}_p(z)$ for the 
BdG wavefunction inside the barrier to distinguish it from that outside 
the barrier. The continuity of the particle current imposes 
usual boundary conditions at the barrier walls, reading
\begin{eqnarray}
  & 
  {\tilde\psi}_p(0)=\psi_p(0),\,
  {\tilde\psi}_p(a)=\psi_p(a),
  \label{B-bound}&\\
  &
  \partial_z{\tilde\psi}_p(0)=
   \partial_z\psi_p(0),\,
  \partial_z{\tilde\psi}_p(a)=
  \partial_z\psi_p(a).&
\label{B-derbound}
\end{eqnarray}
The solution inside the barrier satisfying the boundary condition
(\ref{B-bound}) is 
$
{\tilde\psi}_p(z)
=
\frac{\sinh\kappa_0(a-z)}{\sinh \kappa_0 a}\,\psi_p(0) 
+
\frac{\sinh\kappa_0 z}{\sinh \kappa_0 a}\,\psi_p(a).$
%
Inserting it into the boundary conditions (\ref{B-derbound}) 
for the derivatives, we have:
\begin{eqnarray}
&&
\partial_z\psi_p(0)
+\kappa\,\psi_p(0) 
=
\kappa_t\,\psi_p(a),  
\label{boundN}\\
&&
\partial_z\psi_p(a)
-\kappa\,\psi_p(a)
=-\kappa_t\,\psi_p(0),
\label{boundS}
\end{eqnarray}
where $\kappa=\kappa_0\,{\rm cotanh}\,\kappa_0 a $ and 
$\kappa_t=\kappa_0/\sinh\kappa_0 a$.
Equations (\ref{boundN}) and (\ref{boundS}) serve now as effective 
boundary conditions for the 
BdG equations in the superconductor and the normal system. In 
the limit $\sinh \kappa_0 a\to\infty$, the coupling 
between the "normal" and the "superconducting" functions vanishes, which 
is described by Eqs. (\ref{boundN}) and (\ref{boundS}) 
with zero right-hand sides. 

We use boundary conditions (\ref{boundN}) and 
(\ref{boundS}) to describe Andreev reflection 
at the superconductor-Q1DES interface under
the assumption that the influence of the Q1DES on the 
superconductor can be neglected. 
To proceed, it is convenient to include the
boundary condition (\ref{boundS}) into the BdG equation (\ref{EqS})
by introducing appropriate delta-function terms as follows
\begin{eqnarray}
&
\left[
\epsilon\sigma_3+
\left(E_S+\frac{\hbar^2}{2m}\partial_z^2+{\hat 
U}_S(z)\right)\sigma_0-\Delta(z)i\sigma_2
\right]\psi_p(z)=&
\nonumber\\
&
=-\frac{\kappa_t\hbar^2}{2m}\delta(z-a)\psi_p(0).
\label{EqSmod}
\end{eqnarray}
We note that the admitted singular potential
${\hat U}_S(z) \equiv \frac{\hbar^2}{2m}\delta(z-a)(\partial_z-\kappa)$ 
reproduces Eq.~\ref{boundS} with zero right-hand side ("isolated 
superconductor"). 

The penetration of Andreev bound states into the superconductor at low energies 
is described by a particular solution of
Eq.~(\ref{EqSmod}) generated by the right-hand side containing
the "normal" function $\psi_p(0)$. It can be expressed in terms of 
the matrix Green function 
of Eq.~(\ref{EqSmod}) whose matrix elements are constructed from 
the quasiparticle
$G_{p,\epsilon}(z,z^\prime)$ and condensate (Gorkov's) 
$F_{p,\epsilon}(z,z^\prime)$ Green functions, namely,
\begin{eqnarray}
\psi_p(z)
=
-\frac{\kappa_t\hbar^2}{2m}
\left(
\begin{array}{cc}
G_{p,\epsilon}(z,a) & 
-F_{-p,-\epsilon}(z,a)\\
F_{p,\epsilon}(z,a) &
G_{-p,-\epsilon}(z,a)
\end{array}
\right)
\psi_p(0).\,
\label{sol}
\end{eqnarray}
Here the Green functions satisfy boundary condition (\ref{boundS}) 
with zero right-hand side. Inserting this solution into the boundary 
condition (\ref{boundN}) at the "normal" side and 
neglecting both energy and momentum dependence 
of the Green functions under conditions $|\epsilon|\ll\Delta$ and 
$p\leq p_N\ll p_S$, one finds  
\begin{eqnarray}
\partial_z\psi_p(0)+\kappa\,\psi_p(0)
=\frac{\kappa^2_t\hbar^2}{2m}Fi\sigma_2\psi_p(0),
\label{b}
\end{eqnarray}
where $F\equiv F_{p=\epsilon=0}(a,a)$ is the condensate Green function taken at the boundary of the 
superconductor. We have omitted the terms proportional to $G$
since for $|\epsilon|\ll\Delta$ they would only result in a shift 
of the dispersion.

The right-hand side of the boundary condition (\ref{b}), which is off-diagonal in the 
particle-hole space, takes into account the conversion of a particle
into a hole (and vice versa) due to Andreev reflection, that occurs
simultaneously with normal scattering. In a narrow quantum wire, whose 
thickness is of order of the Fermi wavelength, the anomalous term in the boundary condition 
(\ref{b}) gives rise to an effective pairing energy between particles and holes in the wire.   
Indeed, combining the equation of motion (\ref{EqN}) and the boundary condition (\ref{b}), 
one can write 

\begin{eqnarray}
&
\left[
\epsilon\sigma_3+
\left(E_{N}+\frac{\hbar^2\partial_z^2}{2m}-U_c(z)+{\hat 
U}_N(z)\right)\sigma_0\right]\psi_p(z)=
&\nonumber\\
&
=-\delta(z)\left(\frac{\kappa_t\hbar^2}{2m}\right)^2F
i\sigma_2\psi_p(z),&
\label{Eq}
\end{eqnarray}
where the singular potential ${\hat 
U}_N(z) \equiv -\frac{\hbar^2}{2m}\delta(z)(\partial_z+\kappa)$
is equivalent to the boundary condition (\ref{b}) with zero right-hand 
side. For a weakly coupled Q1DES, the spatial dependence of the BdG 
function $\psi_p(z)\approx\psi_p\,\phi(z)$ is almost 
unaffected by tunneling. Therefore, multiplying Eq.~\ref{Eq} by 
$\phi(z)$ and integrating over $z$, one obtains the following 
one-dimensional equation:
\begin{eqnarray}
&
\left[
\epsilon\sigma_3+
\left(\frac{p^2_F-p^2}{2m}\right)\sigma_0
-E_g i\sigma_2
\right]
\psi_p=0,&
\label{Eq2D}\\
&
E_g \equiv \left(\kappa_t\hbar^2\phi(0)/2m\right)^2F.
&
\label{Eg}
\end{eqnarray}
$E_gi\sigma_2$ {\it plays the role of the effective singlet pairing energy in
the wire}; $p_F$ denotes the Fermi momentum in the Q1DES.

According to Eq.~(\ref{Eq2D}), the excitation spectrum in the Q1DES
is $\epsilon_p^\pm=\pm[v_F^2(|p|-p_F)^2+E_g^{2}]^{1/2}$ 
with the Fermi velocity $v_F=p_F/m$. It has an 
energy gap given by Eq.~(\ref{Eg}) and, hence, the DOS of the normal 
system displays a BCS-like singularity at $E_g$. To estimate $E_g$, 
one can use the condensate Green function of a superconductor 
with a homogeneous pairing potential $\Delta$ at zero energy and parallel momentum,
$F\approx 
W_S^{-1}\sum\nolimits_{p_{z}}\Delta/(\Delta^2+v_S^2(p_z-p_S)^2)$, 
where $v_S=p_S/m$. The integration over $p_z$ gives $F\approx 1/\hbar 
v_S$. The boundary value $\phi(0)$ of the transverse function 
can be estimated using the unperturbed boundary condition 
$\phi(0)=-\kappa^{-1}\partial_z\phi(0)$,
where on the right-hand side one can use the "hard wall" wavefunction
$\phi(z)=(2/W_N)^{1/2}\sin\pi z/W_N$,
which gives $|\phi(0)|\approx\kappa^{-1}(2/W_N)^{1/2}(\pi/W_N)$.
Thus, the effective pairing energy is  
\begin{eqnarray}
E_g=\frac{\hbar}{W_Np_S}\,\frac{1}{\sinh^2 \kappa_0 a}\, E_0,  
\label{Eff_est}
\end{eqnarray}
with $E_0=\hbar^2\pi^2/2mW_N^2$ being the energy of the lowest occupied subband
in the quantum well. Equation (\ref{Eff_est}) is equivalent to 
the one obtained in Ref.~\onlinecite{MiniTh} for a strong delta-shaped barrier.  

Equation (\ref{Eg}) for the minigap $E_g$ provides a link to the 
numerical approach and results discussed earlier.
According to Eq.~(\ref{Eg}), the size of the minigap depends on
the parameter $\kappa_t\hbar^2/2m$ that
characterizes "hopping" between the systems (see Eq.~(\ref{sol})). 
This parameter represents a direct analogue of the coupling constant $\gamma_{NS}$ 
that determines the size of the minigap in the DOS in our numerical study
[see Fig.~\ref{fig:3}(b)].        
Since $E_g$ is quadratic in $\kappa_t\hbar^2/2m$, the numerical value 
of the minigap should scale with $\gamma_{NS}$ as 
\begin{eqnarray}
E_g\propto\gamma^2_{NS},
\label{Eff_num}
\end{eqnarray}
which can indeed be verified numerically (see inset in Fig.~\ref{fig:3}(b)). 
In the next section we will see that the parabolic dependence of the effective 
pairing energy (\ref{Eff_num}) on $\gamma_{NS}$ 
can also be extracted from calculations of the Andreev scattering probability.

\section{Quasiparticle scattering: length and energy dependence} 
\label{sec:S3}

In this section we discuss electron scattering properties that
can be used as an independent and more complete probe of the proximity
effect in finite length parallel N/S contacts. For definiteness we consider
particles in the left region (I) of Fig.~\ref{fig:1} propagating to the right.
When incident at the boundary with the proximity region, these may be:
(a) Andreev reflected, namely, converted into outgoing holes with
the probability $R_A$, (b) normally reflected as outgoing particles, 
i.e., without Andreev conversion, with the probability $R_N$, and (c)
normally transmitted as particles with the probability $T_N$ 
either in the region (II) of Fig. \ref{fig:1}(a) or in the right region (I)
of Fig. \ref{fig:1}(b). Finally, the probability of being
Andreev transmitted to the right is determined via particle conservation,
namely, $1-R_A-R_N-T_N$. In our calculations, this is ensured by the
unitarity of the scattering matrix.

\begin{figure}[t]
\centerline{
\epsfig{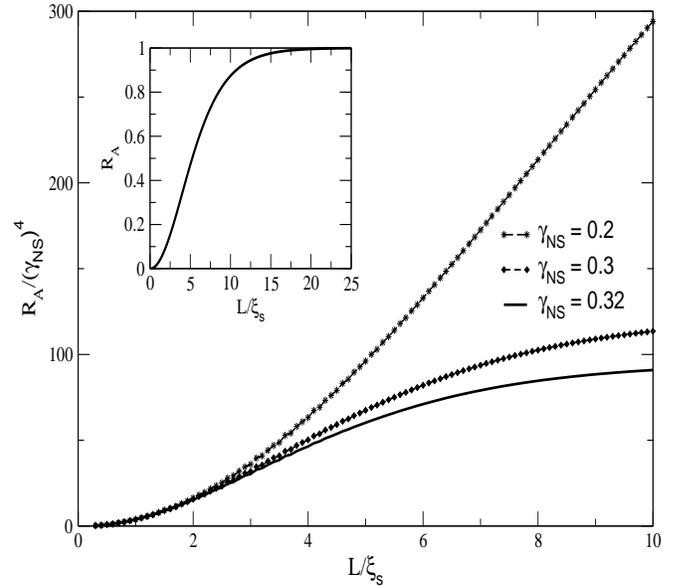}
}
\caption{Andreev reflection probability for various N/S coupling 
constants as a function of the length of the proximity region (II)
in Fig. \ref{fig:2}(b). $R_A$ scales with the interfacial transparency
and at short lengths is quadratic in $L$. For $L$ much larger than the 
proximity-induced coherence length $\xi_N\approx 5\xi_S$, the probability
$R_A$ reaches unity as a manifestation of the reflectionless tunneling
despite the low interfacial transparency
(see inset for $\gamma_{NS}=0.2$).}
\label{fig:4}
\end{figure}

We examine first the dependence of the zero-energy Andreev reflection coefficient
$R_A(\epsilon=0)$ on the length $L$ of the proximity region (II) in
Fig. \ref{fig:1}(b) for different values of the coupling parameter
$\gamma_{NS}$ and without any potential mismatch at the (I)/(II) boundary
($U_o=0$). In conventional N/S/N structures the Andreev coefficient is known 
to scale as $R_A\sim (\Delta\, L)^2$ for $L$ much shorter than the coherence
length $\xi_S$~\cite{PRB95CHL}. According to the results of the previous section,
in our case the effective pairing energy $E_g$ [Eq.~(\ref{Eg})]
should act as $\Delta$ and therefore we expect that $R_A\sim (E_g L)^2$ or,
according to Eq.~\ref{Eff_num}, $R_A \sim \gamma_{NS}^4 L^2$ 
for short enough contacts. This scaling is demonstrated in Fig.~\ref{fig:4} by the
convergence of the appropriately normalized $R_A$ curves
and their parabolic shape at short lengths. As shown in the inset, there is 
a characteristic length $\xi_N\approx 5\xi_S$ beyond which the Andreev probability
$R_A(L)$ approaches its unit limit. Moreover, the ratio of $\xi_N/\xi_S$
coincides with the ratio of the gaps $\Delta/E_g\approx 5$  found from the
analysis of the DOS in the previous section: 
\begin{eqnarray}
\xi_N/\xi_S=\Delta/E_g.
\label{xi_S}
\end{eqnarray}
The overall length dependence implies that the reflectionless tunneling 
builds up due to the strong mixing of particles and holes in long channels. 
In particular, the semiclassical approaches of
Ref.~\onlinecite{vanWees} and \onlinecite{Imry} interpret reflectionless tunneling
in terms of the increase in the cumulative Andreev reflection probability with
increasing number of single Andreev reflections at the N/S boundary in the
limiting case of an infinitely long interface $L/\xi_N\to\infty$.

\begin{figure}[t]
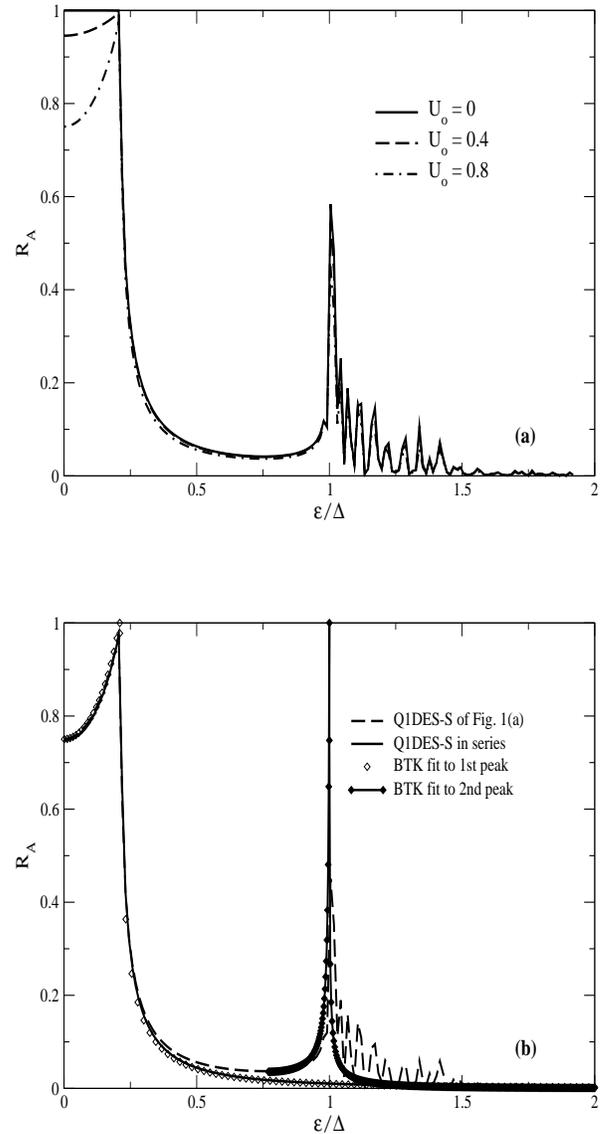

\centerline{
\epsfig{figure=fig5a.eps, width=.9 \hsize ,height=.8 \hsize}
}
\vspace{1.2cm}
\centerline{
\epsfig{figure=fig5b.eps, width=.9 \hsize ,height=.8 \hsize}
}
\caption{
(a) Andreev reflection coefficient for the geometry of
Fig. \ref{fig:1}(a) with $\gamma_{NS}=0.32$
and various values of the potential $U_o$ at (I)/(II) boundary.
For $U_o=0$ the Andreev probability is exactly 1 for energies below $E_g$.
(b) To fit the low- and high-energy peaks we use the formulas of 
the BTK model~\cite{BTK} with parameters 
$\Delta_{BTK}=E_g$, $Z=0.278$ and $\Delta_{BTK}=\Delta$, $Z=28$, respectively.}
\label{fig:5}
\end{figure}

\begin{figure}[t]
\centerline{
\epsfig{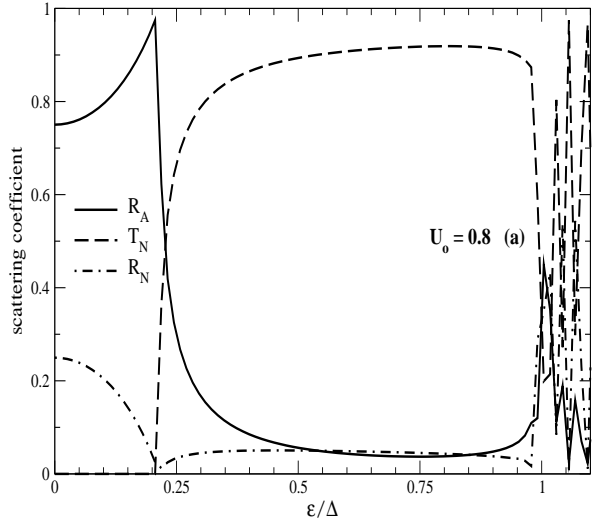}
}
\vspace{1.2cm}
\centerline{
\epsfig{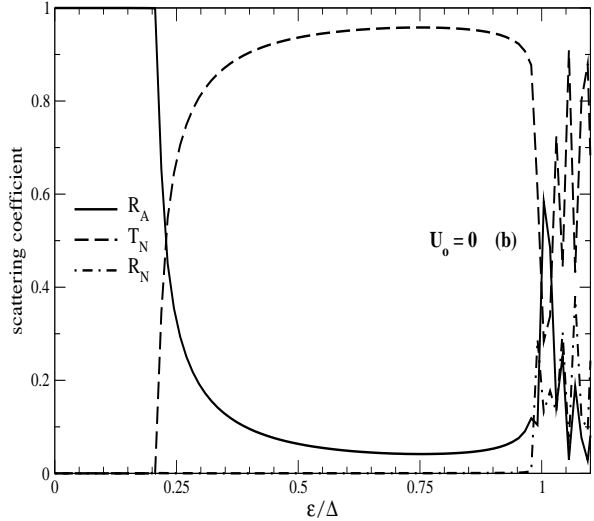}
}
\caption{Andreev reflection $R_A$, normal transmission $T_N$, and 
normal reflection $R_N$ coefficients for the geometry of
Fig. \ref{fig:1}(a) with $\gamma_{NS}=0.32$ and
various potential landscapes in region (II).}
\label{fig:6}
\end{figure}

We now turn to the discussion of scattering of finite-energy quasiparticles
in the semi-infinite geometry of the proximity region [Fig.~\ref{fig:1}(a)]
where the reflectionless tunneling is most pronounced. 
We also take into account a finite potential step $U_o$ at 
the boundary between the normal (I) and proximity (II) regions
that, as has been already mentioned, may arise from the
modification of the Fermi energy in region (II) due to the
coupling to the superconductor. The energy dependence of the Andreev
reflection coefficient is plotted in Fig.~\ref{fig:5}(a) 
for various $U_o$ and $\gamma_{NS}=0.32$. At low energies
$\epsilon\leq E_g=0.21\Delta$ the shape of the dependence $R_A(\epsilon)$ resembles 
that of high-transparency N/S point contacts discussed 
by Blonder, Tinkham and Klapwijk (BTK)~\cite{BTK}.
If there is no potential step $U_o$ between the normal (I) and
proximity (II) regions, the probability $R_A$ equals unity and starts 
to drop at $\epsilon\geq E_g$.
For $U_o\not =0$, finite normal reflection $R_N$ builds up
(shown in Fig. \ref{fig:6}(a)) which results in smaller zero-energy values of $R_A$. 
The appearance of the second narrow peak at $\epsilon = \Delta$~\cite{comment1}
manifests the crossover from the reflectionless tunneling regime,
which involves a two-particle process, to the usual independent electron tunneling
through the barrier.

In Fig.~\ref{fig:5}(b) we demonstrate that the low- and high-energy 
peaks in the dependence of $R_A$ can be independently fitted by the BTK 
model~\cite{BTK}. To fit the low-energy behaviour
we use the formulas of Table II in Ref.~\onlinecite{BTK}
with $\Delta_{BTK} =E_g$ and a small barrier parameter $Z=0.278$.
For the tunneling peak we use the same formulas with $\Delta_{BTK} =\Delta$
and the large barrier parameter $Z=28$. 
In either of the above limiting cases, the fit is almost perfect. 
To describe the crossover between them, a more general 
analytical model is needed.

In Fig.~\ref{fig:6} all non-vanishing scattering coefficients are
plotted for: (a) $U_o=0.8$ and (b) no barrier between
the normal (I) and proximity (II) regions. For $U_o \not= 0$
there is normal reflection of particles at the (I)/(II) interface
caused by the potential mismatch. For $U_o = 0$ it vanishes not only 
below the minigap $\epsilon <E_g$ but also at the intermediate energies
$E_g<\epsilon <\Delta$. This is due to the specific geometry 
of our tunnel junction where quasiparticles with intermediate energies
are mainly transmitted through the channel  
experiencing low-probability Andreev reflection 
(cf the behaviour of $T_N(\epsilon)$ and $R_A(\epsilon)$).  
At higher energies $\epsilon>\Delta$, when the superconductor becomes transparent
for quasiparticles, the lack of the translational invariance of our system 
causes considerable normal scattering and oscillations of all the coefficients
due to the finite thickness of the superconductor.  

\section{Concluding remarks}
\label{sec:S4}

\begin{figure}[t]
\centerline{
\epsfig{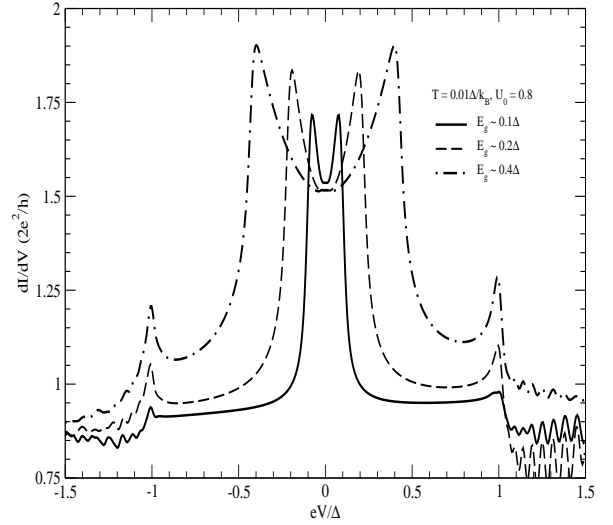}
}
\caption{Differential conductance of the N/S hybrid in Fig.~\ref{fig:5}(a)
at ultralow temperature $T \ll E_g/k_B$.}
\label{fig:7}
\end{figure}

Scattering coefficients are related to the two-, three-, and in general
multi- probe conductances of N/S systems
(see, e.g., Refs.~\onlinecite{BTK},~\onlinecite{Lambert}).
Therefore, the proximity effect discussed in the previous section
should be observable in measurements of the current-voltage characteristics
of such hybrids. In order to explore this possibility and to elaborate
on our discussion of the experimental
reports~\cite{Morpurgo,Kroemer,Weiss,Rahman},
we conclude by focusing on the two-probe differential conductance
$g(eV) \equiv dI/dV$ of the semi-infinite geometry of
Fig. \ref{fig:1}(a). This is given
by \cite{Lambert}
\begin{eqnarray}
g(eV) = \frac{2e^2}{h}
\int_0^\infty d \epsilon \:
\left\{ -\frac{\partial f^p}{\partial \epsilon}\, (1-R_N^p+R_A^p)
\right.
\nonumber\\
\left.
-\frac{\partial f^h}{\partial \epsilon}\, (1-R_N^h+R_A^h)
\right\},
\label{CReq1}
\end{eqnarray}
where $f^{p(h)}=\{\exp[ (\epsilon \mp eV) / k_B T] + 1 \}^{-1}$
with ($-$) for (p)articles and ($+$) for (h)oles. The bias energy $eV$ is
introduced as the difference between the chemical potentials in the normal region
and in the hybrid part of the junction, with the latter taken as reference.

At zero-temperature, Eq.~(\ref{CReq1}) reduces to $g(eV)=1+R_A^p(eV)-R_N^p(eV)=
1+R_A^h(-eV)-R_N^h(-eV)$. Hence, for a small barrier between the normal
(I) and proximity (II) regions of the wire, the dependence of $g(eV)$ at
ultralow $T$ reflects mainly the energy dependence of the Andreev probability
[Fig.~\ref{fig:5}(a)]. The same is true for the length dependence of the zero-bias
conductance. In Fig.~\ref{fig:7} we plot the differential conductance 
for several values of interfacial coupling $\gamma_{NS}$
(i.e., $E_g$) at $T=0.01\Delta/k_B$. Unlike the tunneling peaks at $\pm\Delta/e$,
the proximity-induced anomalies at the minigap energy $\pm E_g/e$ exhibit a strong 
dependence on $\gamma_{NS}$ (cf Fig.~\ref{fig:3}(b)).

\begin{figure}[t]
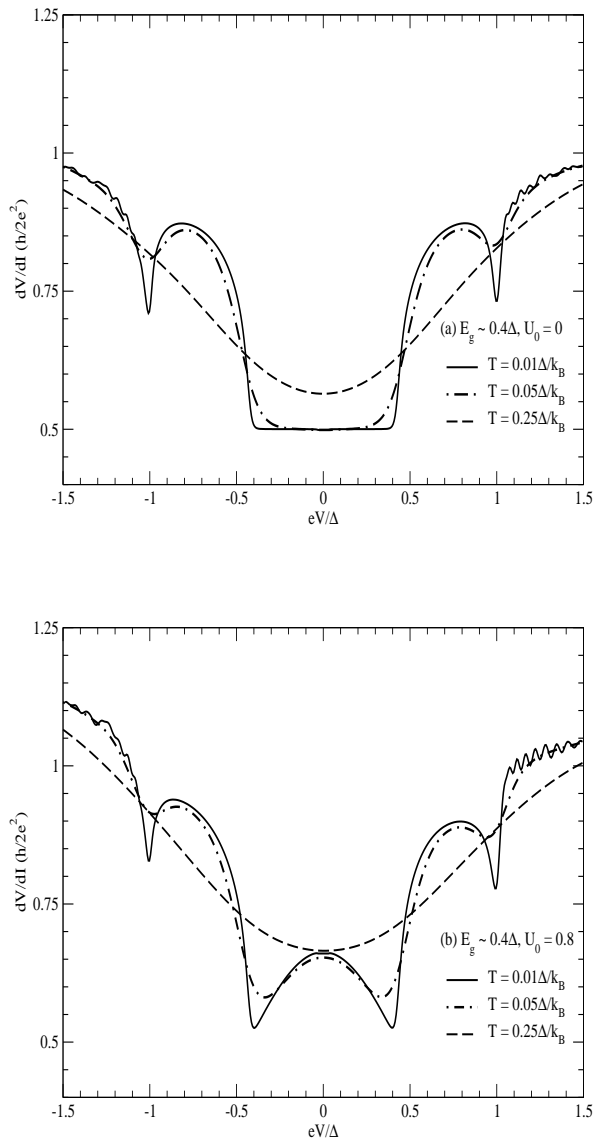

\centerline{
\epsfig{figure=fig8a.eps, width=.9 \hsize ,height=.8 \hsize}
}
\vspace{1.2cm}
\centerline{
\epsfig{figure=fig8b.eps, width=.9 \hsize ,height=.8 \hsize}
}
\caption{Evolution of the differential resistance of the N/S hybrid in
Fig.~\ref{fig:5}(a), as a function of temperature from $k_B T\sim E_g < \Delta$
to $T \ll E_g$ for (a) vanishing and (b) finite potential step between
regions (I) and (II).}
\label{fig:8}
\end{figure}

In Fig.~\ref{fig:8}, the evolution of the differential resistance,
which is defined as the inverse of Eq.~\ref{CReq1}, is shown as a function
of temperature. At intermediate $T \sim E_g/k_B$, features at the scale of the
minigap are smeared and the resistance exhibits an overall dip as a result
of the reflectionless tunnelling.
With decreasing the temperature to $T \ll E_g/k_B$ the resistance curve
develops a finer structure reflecting the energy dependence
of the Andreev reflection probability. For vanishing potential step $U_0$
(Fig.~\ref{fig:8}(a)), there are two minima symmetric to zero bias at the
energies of the superconducting gap. In addition to those,
for $U_0 \not= 0$ (Fig.~\ref{fig:8}(b)) the finite normal reflection 
at $|\epsilon| < E_g$ leads to a zero-bias resistance peak superimposed 
on the Andreev dip.

According to Fig.~\ref{fig:4}, for pronounced Andreev reflection 
the contact length $L$ must be of order of the
proximity-induced coherence length $\xi_N =5\xi_S$. For
$0.01\mu m<\xi_S<0.04\mu m$  typical for Nb electrods,
$\xi_N$ can be estimated as $0.05\mu m<\xi_N<0.2\mu m$. 
On the other hand, in the device of Ref.~\onlinecite{Morpurgo}
the contact overlap between the nanotube and the superconductor
was rather large, about $1\mu m$. Therefore, the condition 
$L>\xi_N$ could be met leading to the observed zero-bias reduction 
of the contact resistance. Another feature of our 
$dV/dI(V)$ curves, namely the appearance of a small 
zero-bias peak superimposed on the Andreev dip due to finite normal scattering at very low $T$ 
(Fig.~\ref{fig:8}(b)), is also consistent with the experimental findings.

We note that in the experiment of Ref.~\onlinecite{Morpurgo} 
the low-bias behaviour of the resistance was sensitive to a gate voltage 
applied to the carbon nanotube. In our model the effect of
the gate voltage can be incorporated into the difference $U_o$ 
between the Fermi energies in the normal (I) 
and proximity (II) regions in Fig.~\ref{fig:1}. 
We have focused on the most interesting case of relatively small $U_o$ when 
normal scattering does not impede the conversion of the quasiparticle current 
into the supercurrent in the proximity region.  
As $U_o$ increases, the low-bias resistance dip in Fig.~\ref{fig:8}(b) eventually 
evolves into an overall peak above the normal state value  
like in non-ideal N/S point contacts~\cite{BTK,PRBNFJ99}.

We emphasize that the zero-bias anomaly discussed here is a property of a
single parallel N/S contact. Although as argued in Ref.~\onlinecite{Morpurgo}
in their S/CN/S devices the two CN/S interfaces acted independently, the role
of the interelectrode coupling remains unclear.
Such a question has been investigated numerically in Ref.~\onlinecite{Wei}
for a somewhat simpler system where a carbon nanotube is connected to a
normal metal and a superconductor via tunnel barriers (N/CN/S).
It was shown that resonant tunneling through Andreev levels in the nanotube
can significantly increase the low-bias subgap conductance similar to the
situation  in mesoscopic N/quantum dot/S structures~\cite{Been}. For a 
comprehensive theory of transport in S/CN/S hybrids this aspect together
with the contact geometry and electron interaction effects must be taken
into account. In addition, one should bear in mind that in the experimental
realizations a number of CNs have been contacted in parallel.

In Refs.~\onlinecite{Kroemer} and~\onlinecite{Weiss}
a strong zero-bias suppression of the resistance was found
in ballistic 2D electron systems in extended planar coupling
to superconductors at $T < \Delta/k_B$. These systems can be considered as a
generalization of that shown in Fig.~\ref{fig:1}. For perfect planar
interfaces, individual channels with possibly different barriers
and interfacial transparencies~\cite{PRBNFJ99} add up independently.
However, interchannel mixing must be considered for rough surfaces.
The same applies when considering experiments in quantum wires~\cite{Rahman}
with few propagating modes. It is worth noting that in this case
the behaviour similar to the low-$T$ differential conductance
of Fig.~\ref{fig:7} was observed.
We believe that the transport anomalies observed in 1D~\cite{Morpurgo},
quasi-1D~\cite{Rahman} and 2D~\cite{Kroemer,Weiss} systems have a contribution of 
a common nature stemming from the proximity-induced mixing of particles and holes
which mediate the conversion of a normal current into a supercurrent along
the contact on the scale of the coherence length $\xi_N=\xi_S\Delta/E_g$
and at energies smaller than the minigap $E_g<\Delta$.

\begin{acknowledgments}
We thank A.F. Morpurgo for communicating the geometrical parameters of
S/CN/S structures studied in Ref.~\onlinecite{Morpurgo}. GF acknowledges
funding by the ATOM CAD project within the {\it Sixth Framework Programme}
of the EU and by the Science Foundation Ireland. This work was also
supported by the  Deutsche Forschungsgemeinschaft (Forschergruppe 370,
Graduiertenkolleg 638).
\end{acknowledgments}



\begin{references}


\bibitem{Fazio}
R. Fazio, F.W.J. Hekking, and A.A. Odintsov, Phys. Rev. Lett. {\bf 74}, 1843 (1995);
C. Winkelholz, R. Fazio, F.W.J. Hekking, and G. Sch\"{o}n,
Phys. Rev. Lett. {\bf 77}, 3200 (1996). 

\bibitem{Maslov}
D.L. Maslov, M. Stone, P.M. Goldbart, and D. Loss,
Phys. Rev. B {\bf 53}, 1548-1557 (1996).
 
\bibitem{Takane}
Y. Takane, J. Phys. Soc. Jpn. {\bf 66}, 537 (1997).

\bibitem{Affleck}
I. Affleck, J.-S. Caux, and A.M. Zagoskin,
Phys. Rev. B {\bf 62}, 1433 (2000).

\bibitem{Wei}
Y. Wei, J. Wang, H. Guo, H. Mehrez, and C. Roland, Phys. Rev. B {\bf 63}, 195412 (2001).

\bibitem{Titov}
M. Titov, N.A. Mortensen, H. Schomerus, and C.W.J. Beenakker, 
Phys. Rev. B {\bf 64}, 134206 (2001).

\bibitem{Gonzalez}
J. Gonzalez, Phys. Rev. Lett. {\bf 87}, 136401 (2001); 
J. Phys.: Condens. Matter {\bf 15}, S2473 (2003).

\bibitem{Vish}
S. Vishveshwara, C. Bena, L. Balents, and M.P.A. Fisher,
Phys. Rev. B {\bf 66}, 165411 (2002).

\bibitem{Nikolic}
B.K. Nikolic, J.K. Freericks, and P. Miller,
Phys. Rev. Lett. 88, 077002 (2002).

\bibitem{Lee}
H.-W. Lee, Hyun C. Lee, Hangmo Yi, and Han-Yong Choi,  Phys. Rev. Lett. {\bf 90}, 247001 (2003).

\bibitem{Jiang}
J. Jiang, L. Yang, J. Dong, and D.Y. Xing,
Phys. Rev. B {\bf 68}, 054519 (2003).

\bibitem{Vecino}
E. Vecino, A. Mart\'{i}n-Rodero, and A.L. Yeyati,
Phys. Rev. B {\bf 68}, 035105 (2003).

\bibitem{Tanaka}
Y. Tanaka, A.A. Golubov, and S. Kashiwaya
Phys. Rev. B {\bf 68}, 054513 (2003).



\bibitem{Kasumov}
A.Yu. Kasumov, R. DeBlock, M. Kociak, B. Reulet, H. Bouchiat, I.I. Khodos, Yu.B. Gorbatov, V.T. Volkov, C. Journet, and M. Burghard, Science {\bf 84}, 1508 (1999);
A. Kasumov, M. Kociak, M. Ferrier, R. DeBlock, S. Gueron, B. Reulet, I. Khodos, O. Stephan, and H. Bouchiat, 
Phys. Rev. B {\bf 68}, 214521 (2003).

\bibitem{Morpurgo}
A.F. Morpurgo, J. Kong, C.M. Marcus, and H. Dai, Science {\bf 286}, 263 (1999);
Physica B {\bf 280} 382 (2000).

\bibitem{Schoenen} 
M.R. Buitelaar, T. Nussbaumer, and C. Sch\"{o}nenberger, Phys. Rev. Lett. {\bf 89}, 256801 (2002).

\bibitem{Haruyama}
J. Haruyama, K. Takazawa, S. Miyadai, A. Takeda, N. Hori, I. Takesue, Y. Kanda, N. Sugiyama, T. Akazaki, 
and H. Takayanagi,  Phys. Rev. B {\bf 68}, 165420 (2003);
J. Haruyama, A. Tokita, N. Kobayashi, M. Nomura, S. Miyadai, K. Takazawa, A. Takeda, and Y. Kanda,
Appl. Phys. Lett. {\bf 84}, 4714 (2004). 


\bibitem{Bockrath}
M. Bockrath, D.H. Cobden, J. Lu, A.G. Rinzler, R.E. Smalley, L. Balents, and P.L. McEuen, Nature {\bf 397}, 598 (1999).

\bibitem{Dres}
M.S. Dresselhaus, G. Dresselhaus, and P.C. Eklund, {\em Science of Fullerenes and Carbon Nanotubes} 
(Academic, San Diego, 1996).

\bibitem{Andreev}
A.F. Andreev,  Zh.\ Exp.\ Teor.\ Fiz.\ {\bf 46}, 1823 (1964)
[Sov.\ Phys.\ JETP {\bf 19}, 1228 (1964)].

\bibitem{BTK}
G.E. Blonder, M. Tinkham, and T.M. Klapwijk,
Phys. Rev. B {\bf 25}, 4515 (1982).



\bibitem{Kroemer}
C. Nguyen, H. Kroemer, and E.L. Hu, Phys. Rev. Lett. {\bf 69}, 2847 (1992).

\bibitem{Rahman}
F. Rahman and T.J. Thornton, Superlat. Microstr., {\bf  25} 767 (1999).

\bibitem{Weiss}
J. Eroms, M. Tolkiehn, D. Weiss, U. R\"{o}ssler, J. DeBoeck, S. Borghs,
Europhys. Lett. {\bf 58}, 569 (2002);
J. Eroms, PhD thesis, Universit{\"a}t Regensburg, 2002.

\bibitem{MiniTh}
A.F. Volkov, P.H.C. Magnee, B.J. van Wees, and T.M. Klapwijk,
Physica C {\bf 242}, 261 (1995).


\bibitem{Cast}
A. Kastalsky, A.W. Kleinsasser, L.H. Greene, R. Bhat, F.P. Milliken 
and J. P. Harbison, 
Phys. Rev. Lett. {\bf 67}, 3026 (1991).

\bibitem{vanWees}
B.J. van Wees, P. de Vries, P. Magnee, and T.M. Klapwijk,
Phys. Rev. Lett. {\bf 69}, 510-513 (1992).


\bibitem{Green}
A.F. Volkov, Phys. Lett. A {\bf 174}, 144 (1993);
A.F. Volkov, A.V. Zaitsev, and T.M. Klapwijk, Physica C {\bf 210}, 21 
(1993).

\bibitem{Been}
C.W.J. Beenakker, Phys. Rev. B {\bf 46}, R12841 (1992);
Rev. Mod. Phys. {\bf 69}, 731 (1997).

\bibitem{Marmorkos}
I.K. Marmorkos, C.W.J. Beenakker, and R.A. Jalabert
Phys. Rev. B {\bf 48}, R2811 (1993).

\bibitem{Nitta}
J. Nitta, T. Akazaki, and H. Takayanagi, Phys. Rev. B {\bf 49} R3659 (1994).

\bibitem{Poirier}
W. Poirier, D. Mailly, and M. Sanquer, Phys. Rev. Lett. {\bf 79},  2105 (1997).

\bibitem{Lambert}
C.J. Lambert and R. Raimondi, J. Phys: Condens. Matter {\bf 10}, 901 
(1998). 

\bibitem{Les}
G.B. Lesovik, A.L. Fauchere, and G. Blatter,
Phys. Rev. B {\bf 55}, 3146 (1997).


\bibitem{Imry}
M. Schechter, Y. Imry, and Y. Levinson,
Phys. Rev. B {\bf 64}, 224513 (2001).


\bibitem{Batov}
I.E. Batov, Th. Sch\"{a}pers, A.A. Golubov, and A.V. Ustinov, 
J. Appl. Phys. {\bf 96}, 3366 (2004).

\bibitem{Ferry}
D. Ferry and S.M. Goodnick,  	
{\it Transport in Nanostructures} (Cambridge University Press, 
Cambridge, 2000)


\bibitem{Petrashov}
V.T. Petrashov, V.N. Antonov, P. Delsing, and T. Claeson, Phys. Rev. 
Lett. {\bf 74}, 
5268-5271 (1995); W. Belzig, R. Shaikhaidarov, V.V. Petrashov, and Yu.V. Nazarov,
Phys. Rev. B {\bf 66}, 220505(R) (2002).

\bibitem{Pannetier}
H. Courtois, Ph. Gandit, D. Mailly, and B. Pannetier,
Phys. Rev. Lett. {\bf 76}, 130-133 (1996);
P. Dubos, H. Courtois, O. Buisson, and B. Pannetier,
{\it ibid}. {\bf 87}, 206801 (2001).


\bibitem{Golubov}
A.A. Golubov, E.P. Houwman, J.G. Gijsbertsen, V.M. Krasnov, J. Flokstra, H. Rogalla, and M.Yu. Kupriyanov, Phys. Rev. B {\bf 51},  1073 (1995);
B.A. Aminov, A.A. Golubov, and M.Yu. Kupriyanov, Phys. Rev. B {\bf 53},  365 (1996).


\bibitem{Belzig}
W. Belzig, C. Bruder, and G. Sch\"{o}n, Phys. Rev. B {\bf 54}, 9443 (1996).

\bibitem{Kad}
A. Kadigrobov, L.Y. Gorelik, R.I. Shekhter, M. Jonson, R.Sh. Shaikhaidarov, V.T. Petrashov, P. Delsing, and T. Claeson, Phys. Rev. B {\bf 60}, 14589 (1999). 

\bibitem{Altland}
A. Altland, B.D. Simons, and D. Taras-Semchuk, Adv. Phys. {\bf 49},  321 (2000).

\bibitem{Anthore}
A. Anthore, H. Pothier, and D. Esteve, Phys. Rev. Lett. {\bf 90} (2003) 127001.

\bibitem{Billiards}
J.A. Melsen, P.W. Brouwer, K.M. Frahm, and C.W.J. Beenakker, Europhys. Lett. {\bf 35}, 7 (1996);
H. Schomerus and C.W.J. Beenakker, Phys. Rev. Lett. {\bf 82}, 2951 (1999);
P. Jacquod, H. Schomerus, and C.W.J. Beenakker, Phys. Rev. Lett. {\bf 90},
207004 (2003).

\bibitem{Ihra}
W. Ihra, M. Leadbeater, J.L. Vega, and K. Richter,
Euro. Phys. J. B. {\bf 21}, 425 (2001).

\bibitem{Cserti}
J. Cserti, A. Korm{\'a}nyos, Z. Kaufmann, J. Koltai, and C.J. Lambert,
Phys. Rev. Lett. {\bf 89}, 057001 (2002);
J. Cserti, P. Polinak, G. Palla, U. Z\"{u}licke, and C. J. Lambert
Phys. Rev. B {\bf 69}, 134514 (2004).

\bibitem{PRB84MB}
R.A. Morrow and K.R. Brownstein, Phys. Rev. B {\bf 30}, 678 (1984).

\bibitem{ratio}
Although $\xi_S/\lambda_F\gg 1$ in real superconducting materials, the
actual value of this ratio does not affect our results other than increasing
significantly the computational cost if made much greater than unity.

\bibitem{JMP62L}
P-O. L{\"o}wdin, J. Math. Phys. {\bf 3}, 969 (1962).

\bibitem{PRB99TSL}
F. Taddei, S. Sanvito, and C.J. Lambert, Phys. Rev. B {\bf 63}, 012404 
(2001).

\bibitem{PRB99SLJ}
S. Sanvito, C.J. Lambert, J.H. Jefferson, and A.M. Bratkovsky,
Phys. Rev. B {\bf 59}, 11936 (1999).

\bibitem{PRB04FKL}
G. Fagas, A.G. Kozorezov, C.J. Lambert {\it et al},
Phys. Rev. B {\bf 60}, 6459 (1999).

\bibitem{CPL04FKE}
G. Fagas, A. Kambili, and M. Elstner, Chem. Phys. Lett.
{\bf 389}, 268 (2004).

\bibitem{TIMES}
We use an optimised F90 implementation developed by one of the
authors (GF). It is generalised to include manipulation of
non-orthogonal matrices, needed to tackle arbitrary geometries, and of
possible non-orthogonal basis sets for applications to materials.


\bibitem{Mark}
M. Leadbeater, PhD Thesis, Lancaster University, 1996.


\bibitem{PRB95CHL}
N.R. Claughton, V.C. Hui, and C.J. Lambert,
Phys. Rev. B {\bf 51}, 11635 (1995).


\bibitem{comment1}
We have checked that $R_A(\epsilon)$
approaches unity at both the minigap (when $U_o \not= 0$)
and superconducting gap energies for selected cases. This requires
a fine choice of the energy grid due to the singular behaviour of the
peaks and, hence, increased computational time. Since our results do 
not depend crucially on this, this practice has been avoided.

\bibitem{PRBNFJ99}
N.A. Mortensen, K. Flensberg, and A.-P. Jauho,
Phys. Rev. B {\bf 59}, 10176 (1999)

\end{references}
\end{document}